\documentclass[preprintnumbers,superscriptaddress,pra]{revtex4}
\usepackage{amssymb}
\usepackage{amsmath}
\usepackage{epsfig}
\usepackage{graphicx}
\usepackage{color}

\input{tcilatex}
\begin{document}

\title{Energy transfer and position measurement in quantum mechanics}
\author{J. C. Ye}
\affiliation{School for Theoretical Physics, School of Physics and Electronics, Hunan
University, Changsha 410082, China}
\author{S. Q. Kuang}
\affiliation{School for Theoretical Physics, School of Physics and Electronics, Hunan
University, Changsha 410082, China}
\author{Z. Li}
\affiliation{School for Theoretical Physics, School of Physics and Electronics, Hunan
University, Changsha 410082, China}
\author{S. Dai}
\affiliation{School for Theoretical Physics, School of Physics and Electronics, Hunan
University, Changsha 410082, China}
\author{Q. H. Liu}
\email{quanhuiliu@gmail.com}
\affiliation{School for Theoretical Physics, School of Physics and Electronics, Hunan
University, Changsha 410082, China}
\affiliation{Synergetic Innovation Center for Quantum Effects and Applications (SICQEA),
Hunan Normal University,Changsha 410081, China}
\date{\today }

\begin{abstract}
The Dirac delta function can be defined by the limitation of the rectangular
function covering a unit area with decrease of the width of the rectangle to
zero, and in quantum mechanics the eigenvectors of the position operator
take the form of the delta function. When discussing the position
measurement in quantum mechanics, one is prompted by the mathematical
convention that uses the rectangular wave function of sufficiently narrow
width to approximate the delta function in order to making the state of the
position physical. We argue that such an approximation is improper in
physics, because during the position measurement the energy transfer to the
particle might be infinitely large. The continuous and square-integrable
functions of both sharp peak and sufficiently narrow width can then be
better approximations of the delta function to represent the physical states
of position. When the slit experiment is taken as an apparatus of position
measurement, no matter what potential is used to model the slit, only the
ground state of the slit-dependent wave function matters.
\end{abstract}

\maketitle

\section{Introduction}

The measurement postulate of quantum mechanics can be divided into two
parts. \cite{cohen} The first part is: the only possible result of the
measurement of a physical quantity A is one of the eigenvalue of the
corresponding observable $\hat{A}$. The second is (\textit{case of a
continuous non-degenerate spectrum}): when the physics quantity A is
measured on a system in the normalized state $\left\vert \psi \right\rangle $%
, the probability $dP(\alpha )$\ of obtaining a result included between $%
\alpha $ and $\alpha +d\alpha $ is equal to, 
\begin{equation}
dP(\alpha )=\left\vert \left\langle u_{\alpha }\right\vert \left. \psi
\right\rangle \right\vert ^{2}d\alpha ,  \label{1}
\end{equation}%
where $\left\vert u_{\alpha }\right\rangle $ is the eigenvector
corresponding to the eigenvalue $\alpha $ of the observable $\hat{A}$
associated with A, determined by, 
\begin{equation}
\hat{A}\left\vert u_{\alpha }\right\rangle =\alpha \left\vert u_{\alpha
}\right\rangle .  \label{2}
\end{equation}%
For \textit{the case of a discrete non-degenerate spectrum,} we need to
replace Eq. (\ref{1}) by $P(a_{n})=\left\vert \left\langle u_{n}\right\vert
\left. \psi \right\rangle \right\vert ^{2}$ where $n$ denotes the discrete
index and the $n$-th eigen-value and eigen-state $\left\{ a_{n},\left\vert
u_{n}\right\rangle \right\} $ are determined by $\hat{A}\left\vert
u_{n}\right\rangle =a_{n}\left\vert u_{n}\right\rangle $. In our study, for
simplicity, all operators $\hat{A}$ have non-degenerate spectrum only\textit{%
.}

We apply the measurement postulate to the position measurement of a particle
in a quantum state $\left\vert \psi \right\rangle $, and for simplicity we
assume that this particle moves in a one-dimensional potential. The
eigenvector of position operator $\hat{x}$ takes a delta function $\delta
(x-x^{\prime })$ with eigenvalue $x^{\prime }$, obeying following equation,%
\begin{equation}
\hat{x}\delta (x-x^{\prime })=x^{\prime }\delta (x-x^{\prime }).  \label{3}
\end{equation}%
When the position $x$ is measured in any quantum state $\left\vert \psi
\right\rangle $, no experiment can yield a result with complete accuracy.
Let us assume that the opposite is true, i.e., assume that after the
measurement, the particle will collapse to an eigen-state $\delta
(x-x^{\prime })$ with eigenvalue $\hat{x}^{\prime }$. We can expand this
delta-function in terms of eigen-states of the Hamiltonian $H$ of the
particle, 
\begin{equation}
\delta (x-x^{\prime })=\sum_{n}\varphi _{n}^{\ast }(x^{\prime })\varphi
_{n}(x),  \label{4}
\end{equation}%
where $\varphi _{n}(x)$ is the $n$-th eigen-state of the Hamiltonian $H$
with energy eigenvalue $E_{n}$, i.e., we have, 
\begin{equation}
\hat{H}\varphi _{n}(x)=E_{n}\varphi _{n}(x).  \label{5}
\end{equation}%
Result (\ref{4}) shows that the probability of the particle in an arbitrary
energy eigen-state $\varphi _{n}(x)$ is $\left\vert \varphi _{n}(x^{\prime
})\right\vert ^{2}$. Not only the total probability diverges as $%
\sum_{n}\left\vert \varphi _{n}(x^{\prime })\right\vert ^{2}=\delta
(x^{\prime }-x^{\prime })=\delta (0)\rightarrow \infty $, but also does the
expectation value of the energy as $\sum_{n}\left\vert \varphi
_{n}(x^{\prime })\right\vert ^{2}E_{n}=E_{0}\sum_{n}\left\vert \varphi
_{n}(x^{\prime })\right\vert ^{2}\left( E_{n}/E_{0}\right)
>E_{0}\sum_{n}\left\vert \varphi _{n}(x^{\prime })\right\vert
^{2}\rightarrow \infty $, where the fact that the Hamiltonian operator $\hat{%
H}$ is in general self-adjoint and bounded from below, i.e., $0\leq
E_{0}<E_{1}<E_{2}<...$, is used and the smallest value of the eigenvalues is
assumed not less than zero. These two results are absurd, all due to the
non-normalizability of the delta function. A question arises immediately: If
the normalizable rectangular wave function instead of the delta function
sufficient? If not, which one is proper? The present paper investigates
these problems and those related.

In section II, we will demonstrate that the conventional normalizable
rectangular wave function is unsatisfactory. In section III and IV, we show
how the uncertainty principle imposes a strong requirement on position
measurement. In section V, a brief summary is given.

\section{Energy divergence with rectangular wave functions}

In a realistic experiment of position measurement, the position of the
particle can only be found to be within a sufficiently narrow interval that
nevertheless contains uncountably infinite many position eigenvalues because
of the continuous spectrum of the position, and possibly these eigenvalues
may be distributed around a point that is more probable. The most common
means to get over the divergence problem is to use some approximated
representations of the delta function. The simplest and the conventional
choice is the rectangular function, \cite%
{cohen,weinberg,edu1,edu1-1,edu2,edu3,edu4,edu5} 
\begin{equation}
\chi _{\varepsilon }(x-x_{0})=\left\{ 
\begin{array}{c}
1/\sqrt{\varepsilon } \\ 
0%
\end{array}%
,%
\begin{array}{c}
x\in \left[ x_{0}-\varepsilon /2,x_{0}+\varepsilon /2\right]  \\ 
\text{otherwise}%
\end{array}%
\right. ,\varepsilon >0.  \label{6}
\end{equation}%
This function is normalized to unity, i.e., $\int \left\vert \chi
_{\varepsilon }(x-x_{0})\right\vert ^{2}dx=1$, and also we have, 
\begin{equation}
\lim_{\varepsilon \rightarrow 0^{+}}\int_{x_{0}-\varepsilon
/2}^{x_{0}+\varepsilon /2}\left\vert \chi _{\varepsilon
}(x-x_{0})\right\vert ^{2}f(x)dx=\lim_{\varepsilon \rightarrow
0^{+}}\int_{x_{0}-\varepsilon /2}^{x_{0}+\varepsilon /2}\frac{1}{\varepsilon 
}\left( f(x_{0})+f^{\prime }(x_{0})(x-x_{0})+O((x-x_{0})^{2}\right)
dx=f(x_{0}),  \label{6.1}
\end{equation}%
where $f(x)$ is a continuous function defined in the same interval in which
the wave function under study is defined. This choice of the approximation
of the delta function is still improper. The problem lies in that the
rectangular wave function (\ref{6}) belongs to the differentiability class $%
C^{0}$ rather than the class higher than $C^{1}$. We demonstrate it in the
following.

Let us measure the position of the particle in a one-dimensional infinitely
deep well of width $a$ defined in an interval $x\in \left( 0,a\right) $ and
the position happens to be approximately in the middle $x_{0}=a/2$. The
stationary state energies the states for the particle in the well are,
respectively, 
\begin{equation}
E_{n}=\frac{(n\pi \hbar )^{2}}{2ma^{2}},\varphi _{n}(x)=\sqrt{\frac{2}{a}}%
\sin \left( \frac{n\pi x}{a}\right) ,n=1,2,3,...,  \label{6.2}
\end{equation}%
where $m$ is the mass of the particle. Once the delta function is
approximated by the rectangular wave function (\ref{6}), each value $E_{n}$
can be found with following probability, \cite{cohen}%
\begin{equation}
P(E_{n})=\left\vert \int \varphi _{n}^{\ast }(x)\chi _{\varepsilon }(x-\frac{%
a}{2})dx\right\vert ^{2}=\frac{8a}{\varepsilon }\left\{ 
\begin{array}{c}
\frac{1}{\left( n\pi \right) ^{2}}\sin ^{2}(\frac{n\pi \varepsilon }{2a}) \\ 
0%
\end{array}%
\begin{array}{c}
\text{odd }n \\ 
\text{even }n%
\end{array}%
\right. .  \label{7}
\end{equation}%
It is normalized to unity. However, the energy expectation diverges as, no
matter what value of $\varepsilon $ is given, 
\begin{equation}
\sum_{n}P(E_{n})E_{n}=\sum_{n=1}^{\infty }P(E_{2n-1})E_{2n-1}=\frac{4\hbar
^{2}}{ma\varepsilon }\sum_{n=1}^{\infty }\sin ^{2}\left[ \frac{(2n-1)\pi
\varepsilon }{2a}\right] \rightarrow \infty .  \label{8}
\end{equation}%
It is totally unacceptable for it implies that at the moment of the
measurement, the particle gains an infinite amount of energy. It is
impossible in physics.

It is easily to verify that utilization of the one-dimensional simple
harmonic oscillator, instead of the one-dimensional infinitely deep well,
can give the same divergence (\ref{8}) as well. In fact, at the
discontinuous points of the function (\ref{6}), action of momentum operator
leads to $\delta $-functions in the following way, 
\begin{equation}
p_{x}\chi _{\varepsilon }(x-x_{0})=-\frac{i\hbar }{\sqrt{\varepsilon }}%
\left( \delta (x-x_{0}+\frac{\varepsilon }{2})-\delta (x-x_{0}-\frac{%
\varepsilon }{2})\right) .  \label{9}
\end{equation}%
Thus it is the discontinuity in (\ref{6}) that results in the divergence.

After all, if one likes to use the rectangular wave functions for some
purposes, he must bear in mind that these functions have fatal flaws.

\section{Position measurement and uncertainty principle}

In order to construct proper position function in realistic experiment, we
must resort to continuous and square-integrable function of
differentiability class not less than $C^{1}$ that has both sharp peak and
sufficiently narrow width to approximate the delta function. The form of the
continuous functions does not matter but the sharpness matters for it
characterizes resolution power of the measuring apparatus. The simplest
possible choose is, 
\begin{equation}
\chi _{\varepsilon }(x-x_{0})=\left\{ 
\begin{array}{c}
\sqrt{\frac{2}{\varepsilon }}\sin \left( \frac{\pi \left(
x-x_{0}+\varepsilon /2\right) }{\varepsilon }\right) \\ 
0%
\end{array}%
,%
\begin{array}{c}
x\in \left[ x_{0}-\varepsilon /2,x_{0}+\varepsilon /2\right] \\ 
\text{otherwise}%
\end{array}%
\right. ,\varepsilon >0.  \label{10}
\end{equation}%
Its probability density behaves like a delta function, for we have not only $%
\int \left\vert \chi _{\varepsilon }(x-x_{0})\right\vert ^{2}dx=1$ but also, 
\begin{equation}
\lim_{\varepsilon \rightarrow 0^{+}}\int_{x_{0}-\varepsilon
/2}^{x_{0}+\varepsilon /2}\frac{2}{\varepsilon }\sin ^{2}\left( \frac{\pi
\left( x-x_{0}+\varepsilon /2\right) }{\varepsilon }\right)
f(x)dx=f(x_{0})\int \left\vert \chi _{\varepsilon }(x-x_{0})\right\vert
^{2}dx=f(x_{0}).  \label{11}
\end{equation}%
Now the energy expectation values\ over state $\chi _{\varepsilon }(x-x_{0})$
(\ref{10}) is finite no matter what small values of $\varepsilon $ are
given, and the values are evidently,%
\begin{equation}
\frac{(\pi \hbar )^{2}}{2m\varepsilon ^{2}}\sim \frac{h^{2}}{m\varepsilon
^{2}}\sim \frac{\delta p^{2}}{m},  \label{12}
\end{equation}%
where $\delta p\sim h/\varepsilon $, as indicated by the uncertainty
principle $\delta x\delta p\sim h$\ with $\delta x\sim \varepsilon $.

One can instead take the Gaussian function approximation of the delta
function, rather than (\ref{10}), and the similar results will be found.
Explicitly, we take the following Gaussian function approximation, 
\begin{equation}
\chi _{\varepsilon }(x-x_{0})=\left( \frac{1}{\sqrt{\pi }\varepsilon }%
\right) ^{1/2}\exp \left[ -\frac{\left( x-x_{0}\right) ^{2}}{2\varepsilon
^{2}}\right] ,\varepsilon >0.  \label{13}
\end{equation}%
The energy expectation value is also finite provided that $\varepsilon $
does not vanish, and the value is given by,%
\begin{equation}
\frac{\hbar ^{2}}{4m\varepsilon ^{2}}\sim \frac{h^{2}}{m\varepsilon ^{2}}%
\sim \frac{\delta p^{2}}{m}.  \label{14}
\end{equation}

Above results (\ref{12}) and (\ref{14}) are physically significant in
following aspects: 1. If the particle before the position measurement has
energy much less than $h^{2}/(m\varepsilon ^{2})$, we must adjust $%
\varepsilon $ to be a large value such that the momentum uncertainty $\delta
p\sim h/\varepsilon $ becomes smaller. 2. The measurement apparatus acts as
an energy conversion that at the moment of measurement converts part of the
energy from the particle before measurement to the particle whose position
is under measurement, and the amount of the energy is quite definite as
shown in (\ref{12}) or (\ref{14}). These two points will be further
discussed in the following section. 3. The position measurement must be of a
very complicated process, which is far beyond full understanding. \cite%
{edu1-1,Wise} It seems to us that annihilations and creations happen during
the measurement.

\section{Comments on the single slit experiment as a possible apparatus of
position measurement}

Now we try to take the single slit experiment to measure the $x$-component
position of a particle via detect its $x$-component momentum on the remote
screen (the Fraunhofer diffraction) after passing through a single narrow
slit of finite width. For this propose, the diaphragm is kept in the plane $%
z=0$, a slit of small width $\delta x=a$ is made with center at $x=0$, and
let the slit be straight and very long along the $y$-axis. It is then that
only the component along the $x$-axis of the incident free particle is
affected by diffraction. A monochromatic wave travelling along $z$-axis with
energy $E_{0}=p_{x}^{2}/2m$ and incident on the first screen which contains
a slit. The emerging wave then arrives at the screen to form a diffraction
pattern. The bright bands correspond to interference maxima, and the dark
bands interference minima. Since the zeroth maximum has dominative part of
the probability, the first dark band offers the best characterization of
uncertainty of the momentum $\delta p_{x}$ that is also the momentum gained
along $x$-axis for the particle passing through the slit, 
\begin{equation}
\delta p_{x}\approx \frac{h}{a}  \label{15}
\end{equation}%
This result is compatible with results (\ref{12}) and (\ref{14}). To note
that within the single slit experiment, for a given openness $a$ of the
slit, once the incident waves have energies $E_{0}=p_{x}^{2}/2m$ lower than $%
h^{2}/2ma^{2}$, there would be no quantum state that contains nonvanishing $x
$-dependent part. In other words, the interference maxima are no longer
appreciable. On the other hand, once the incident waves have energies $%
E_{0}=p_{x}^{2}/2m$ lower than $h^{2}/2ma^{2}$, the single slit experiment
is not suitable apparatus of position measurement.

It should be stressed that when above single slit experiment is taken as an
apparatus of position measurement, no matter what potential is used to model
the slit, only the ground state of the $x$-dependent wave function matters.
This is because, with increase of the quantum numbers in the $x$-dependent
stationary states of\ the wave function, the uncertainties $\delta x\delta
p_{x}$ will increase, and the relation (\ref{15}) will be broken. In other
words, the measurement apparatus inevitably causes some disturbances in both
momentum and energy and it is reasonable to require that these disturbances
keep smallest possible.

\section{Conclusions}

The simplest and the conventional choice of taking the rectangular function
as an approximation of delta function, following the mathematical
convention, is improper as it is used to represent a quantum state.
Calculations of this approximation in position measurement show that the
amount of the energy gained at the moment of measurement might be infinitely
large. Therefore, we have to choose the continuous and square-integrable
functions of both sharp peak and sufficiently narrow width as better
approximations of the delta function representing the physical states of
position. The detailed structure of forms of the continuous functions does
not matter but the sharpness matters for it characterizes resolution power
of the measuring apparatus. Similar discussions are applicable to the
approximation of delta function representing the physical states of
momentum, which can be easily done by the readers as an exercise.

Since the single slit experiment is frequently taken as an apparatus of
position measurement, our results indicate that no matter what potential is
used to model the slit, only the ground state of the slit-dependent wave
function matters.

\begin{acknowledgments}
This work is financially supported by National Natural Science Foundation of
China under Grant No. 11675051.
\end{acknowledgments}

\end{document}